# Oppie Op-ed: Reflections on Christopher Nolan's *Oppenheimer*

Michel Janssen[1]

In this article, at the invitation of my University of Minnesota colleague and editor of this newsletter, Oriol Valls, I want to share some thoughts about Christopher Nolan's movie *Oppenheimer* based on the Pulitzer-prize-winning book *American Prometheus: The Triumph and Tragedy of J. Robert Oppenheimer* by Kai Bird and Martin Sherwin. I am by no means an expert on Oppenheimer or the Manhattan project. But as the resident historian in our School of Physics and Astronomy, I was called upon to give a talk preparing our students and faculty for the movie. A recording of my talk was put on a YouTube channel of the university and has gathered over 2,000 hits (the URL is www.youtube.com/watch?v=_TQJEMC6mkk). I gave this talk shortly before the movie was released and I tried to anticipate—on the basis of the book, the cast list, and the trailer—which parts of this complicated story would make it into the movie. That made preparing the talk fun, provided my audience with an insurance policy of sorts against spoilers, but also turned the movie's release date into my talk's expiration date (although several people have told me my talk helped them understand the plot better *after* they saw the movie).[2] Rather than summarize my talk, I figured I would highlight some elements in the movie that I missed in my talk and organize my article around those. Unlike my talk, this article thus calls for a **SPOILER ALERT**: in what follows, I will assume that the reader has seen the movie, which I have meanwhile seen twice and thoroughly enjoyed both times. And it made me want to keep reading about the topic. One of the books I picked up is Gregg Herken's *Brotherhood of the Bomb. The Tangled Lives and Loyalties of Robert Oppenheimer, Ernest Lawrence and Edward Teller* (New York: Henry Holt, 2002). I already thought it was excellent the first time I read it shortly after it came out—I found it much more manageable than Richard Rhodes' classic *The Making of the Atomic Bomb* (New York: Simon & Schuster, 1986)—but feeling, thanks in no small measure to the movie, that I know many of the characters much better now, I liked it even more this time around. If the movie (and the book by Bird and Sherwin) also left you wanting more, you may want to check out Herken's book.

Given that an actor as prominent as Robert Downey Jr. was cast as Lewis Strauss, Oppenheimer's nemesis, it was obvious that the movie would deal as much with the 1954 security hearings as with the Manhattan project, Los Alamos and the Trinity test. However, I completely failed to anticipate the clever way in which the movie handles the conflict between Oppenheimer and Strauss. That was not for a lack of clues in the cast list. The most important one was that Rami Malek—Freddie Mercury in the Queen movie *Bohemian Rhapsody*—was listed as playing David Hill. Why was such a well-known actor playing a character I had never even heard of? Hill is not mentioned anywhere in Bird and Sherwin's book. He now has his own wikipedia page but I'm pretty sure that page did not exist before the movie was released. I was similarly puzzled by the names of several senators in the cast list: McGee, Bartlett, Pastore, Scott … What was their connection to Oppenheimer? I would have expected Senator Brien McMahon, author of the 1946 Atomic Energy Act, which placed all nuclear matters under civilian rather than military control and led to the establishment of the Atomic Energy Commission (AEC). Why McGee and not McMahon?

---

[1] School of Physics and Astronomy, University of Minnesota. Currently Visiting Fellow, Lichtenberg Group for History and Philosophy of Physics, University of Bonn.

[2] I have since given a post-movie version of the talk, based in part on this opinion piece. A recording of it can be found here: https://mediaspace.umn.edu/media/t/1_cyo0o4ed.

The answer to these questions had actually been staring me in the face: on Strauss' wikipedia page I had read about his 1959 Senate confirmation hearings after he'd been nominated by President Eisenhower to be his secretary of commerce. The senators in the movie are the ones involved in these confirmation hearings. Strauss' wikipedia page even mentions that David Hill testified against him but I don't recall seeing that bit before and I suspect it was added after the movie came out. Yet, even if it had been there all along, I doubt I would have recognized the narrative potential of these confirmation hearings. Christopher Nolan did. It was a brilliant move on his part to turn these hearings into one of the movie's central storylines. It is deeply satisfying to see the bad guy—and Robert Downey Jr. deserves an Oscar for the number he does on "Tugboat Admiral" Lewis Strauss!—get his comeuppance in the end. What makes this especially sweet is that Strauss' downfall closely mirrors Oppenheimer's, which Strauss orchestrated with such evil care. It is the result not of a trial, covered by well-established rules of engagement, but of a hearing in which one makes up the rules as one goes along—a "kangaroo court" as Bird and Sherwin, echoing AEC counsel Joe Volpe, call it in Oppenheimer's case. Accordingly, my favorite line in the movie, applied to both hearings, is: "We don't convict, we just deny." It is likewise satisfying to see Oppenheimer's nasty "prosecutor" Roger Robb (portrayed by Jason Clarke) get his comeuppance. A defiant Kitty Oppenheimer (portrayed beautifully by Emily Blunt) shows him up, in sharp contrast to her husband's stoic martyrdom demeanor as he is being grilled by Robb.

Once I realized the importance of Strauss' confirmation hearing, it was easy to find out more about it. Eisenhower called the day Strauss was denied his cabinet post "the second most shameful day in Senate history" (after Andrew Johnson's impeachment trial). Given the way the situation is portrayed in the movie, an article in *Time Magazine* of Monday, May 18, 1959, is surprisingly sympathetic to Strauss: "What was supposed to be a confirmation hearing … turned out last week to be an undisguised inquisition. To begin with, the Senate Interstate and Foreign Commerce Committee took an unusual step in bringing in a special counsel for the hearing. Committee and counsel called only hostile witnesses, gave Strauss no notice of who would be appearing against him. With witnesses day after day pouring personal rancor into the headlines, the weird sessions added up to one of the bitterest attacks on a presidential Cabinet appointee in the nation's history." Much of this applies verbatim to the Oppenheimer hearings five years earlier. The piece also touches on Hill's testimony. Hill, so *Time Magazine*, "accused Strauss of, among other things, distorting truth and usurping authority. Pennsylvania's Republican Senator Hugh Scott [in Nolan's cast list] remarked that Hill's statement was 'extremely well prepared.' Did he get any help in preparing it from 'anyone connected with the Senate or with any Senate Staff member?' An uneasy silence fell. Then the committee's Special Counsel … spoke up: 'The witness discussed several matters with me, Senator Scott.'" I would have liked to see that exchange included in the movie. The attentive viewer will have spotted David Hill/Rami Malek twice before his big scene at the end of the movie. We first meet him when Oppenheimer visits the laboratory at the University of Chicago at Stagg Field Stadium. We see him a second time in Washington DC in 1945, sometime between Germany's and Japan's surrender. This is the scene in which Oppenheimer knocks a clipboard out of his hands, presumably with the petition that the person standing next to him, Leo Szilárd (played by Máté Haumann), wanted to give to Oppenheimer. The signatories of this petition (known as the Franck report after German emigré and Chicago physicist James Franck) were asking for a demonstration rather than combat use of the atomic bomb,

The Strauss confirmation hearing is not the only scene taking place in the Senate in the movie. The movie also covers the Senate hearing about sharing radioisotopes with Norway, in which Oppenheimer ridicules Strauss (with Nolan changing Oppie's "vitamins"-punchline to the even punchier: "but more important than a sandwich"). This hearing took place in June 1949, a week after Oppenheimer's appearance and a day before his brother Frank's appearance before the House Un-American Activities Committee (HUAC), neither of which made it into the movie. That surprised me. This is where Robert threw some of his students under the bus. This scene would thus have helped Nolan underscore the somewhat slippery character of his main protagonist. Nolan leaves it largely to Oppenheimer's enemies, Strauss and Robb, to highlight some of the more dubious aspects of his character. The best example may be the bitter speech by Strauss/Downey Jr. after he is denied his cabinet post in which he argues that Oppenheimer should be thanking him because he, Strauss, helped Oppenheimer become the martyr he always wanted to be. I don't know whether Strauss actually said that, but the sentiment, I'm afraid, cannot easily be dismissed.

As I indicated above, the most important Senate action missing from the movie is the establishment of the McMahon atomic energy act in 1946. Leaving this out also affects the introduction of someone who does play an important role in the movie: William Borden (played by David Dastmalchian). Borden, who wrote the letter with derogatory information about Oppenheimer to the FBI, the letter that set the wheels in motion for the 1954 hearings, was an aide to Senator McMahon. A Democrat from Kentucky, McMahon chaired the congressional Joint Committee on Atomic Energy during the years of the Truman administration. He died a few months before the 1952 elections in which the Republicans took back both the presidency and the Senate. Staffer Borden thus lost his job. Given McMahon's push in 1946 for civilian control of atomic matters, one naturally thinks that he was one of Oppenheimer's allies (and this is certainly the impression I gave in my talk). Which raises the question why his aide had it in for Oppenheimer. Herken's *Brotherhood of the Bomb* answers that question in detail. In the late 1940s and early 1950s, McMahon and Borden were clearly on the side of Strauss and Edward "the real Dr. Strangelove" Teller (a stellar performance by Benny Safdie). Both McMahon and Borden were strongly in favor of the crash program to develop the H-bomb and were extremely annoyed with Oppenheimer's lobby against it. From his hospital bed in 1952, not long before he died, McMahon warned Truman that he would initiate impeachment proceedings against him should Truman decide to postpone an upcoming test of an H-bomb (Herken, p. 256). In the early 1950s, McMahon and Borden were also vehemently opposed to the idea, favored by Oppenheimer, to explore the possibility of a test ban on nuclear weapons with the Soviets. In fact, an important trigger for Borden's letter about Oppenheimer to the FBI is that, early in Eisenhower's presidency, it looked as if Oppenheimer had the president's ear whereas the Strauss camp clearly had gained the upper hand toward the end of the Truman administration. Recall that, in late August 1949, the Soviets exploded an exact copy of "fat man", the plutonium bomb dropped on Nagasaki, and that, in early February 1950, Los Alamos spy Klaus Fuchs confessed that he had transmitted the relevant blueprints to them (some of the Russian scientists who worked on that program told my Minnesota colleague, Misha Shifman, that they made sure to include every screw shown in the drawings even if they had no idea what it was for).

Following Bird and Sherwin, Nolan recognizes these events as a critical turning point in the story. Despite the unanimous recommendation of Oppenheimer's General Advisory Committee (GAC) to the AEC *not* to develop the H-bomb, Truman decided to do so anyway. Nolan took some liberty with the historical record by having the Fuchs bombshell and the news about the Soviet bomb break at the same time but that's certainly defensible: Strauss already knew that the FBI was on to Fuchs when the discussions about how to respond to the Soviet bomb took place (Herken, p. 213). Nolan has a bunch of scientists, politicians and military men (I could not identify them all) sit around a big round table (with a big bouquet of flowers that keeps being moved around for no discernible purpose) to discuss how to respond to the Soviet bomb. Brigadier General Kenneth D. Nichols, played by Dane DeHaan, is one of those present. Colonel Nichols (he was promoted shortly after the war) had been General Groves' right-hand man on the Manhattan project and continued to serve the military and the government in various capacities related to matters concerning atomic weapons. I overlooked the importance of Nichols in my talk (I flashed up the organizational chart of the Manhattan project at one point where Nichols appears just below Groves but only mentioned Nichols toward the end of my talk). Nolan gives Nichols his due. We see him run security, for instance, at Los Alamos. Both Groves and Oppenheimer treat him with disdain (I recall Groves/Damon handing Nichols/DeHaan his coat and telling him to get it dry-cleaned and Nichols/DeHaan snapping at Oppenheimer/Murphy that it is not his fault that it takes so long to get his security clearance). The moviegoer understands that Nichols does not care for Oppenheimer. The revelation about Fuchs during the discussion about how to respond to the Soviet bomb gives Nichols/DeHaan an opportunity to complain about the lax security at Los Alamos. That nicely sets things up for Nichols' role in the 1954 security hearings. That part, however, is not in the movie. In the movie, the hearings end with the 2:1 recommendation of the personnel security board chaired by Gordon Gray to strip Oppenheimer of his security clearance. But that was just a recommendation. It still had to be voted on by the AEC. In 1953, Nichols had become the General Manager of the AEC. Worried that his fellow commissioners might not follow the Gray board's recommendation, Strauss had Nichols add a letter putting their spin on the security board's report. In this highly prejudicial letter, Nichols wrote that "the record showed" that Oppenheimer is "a communist in every respect except … that he did not carry a party card." Since the AEC upheld the "verdict" of the Gray board, I can I understand why Nolan left out this part but, given Nichols' prominent role in the rest of the movie, it would have been appropriate to have him twist the knife that Strauss stuck in Oppenheimer's back.

There is another connection I missed in my talk and I'm grateful to Nolan for making it for me. This has to do with Boris Pash (played by Casey Affleck). Pash is the one to whom Oppenheimer first mentions the Chevalier incident (providing actors Jefferson Hall and Guy Burnet with rewarding parts as the Oppenheimers' friend Haakon Chevalier and the commie busybody George Eltenton, respectively). Oppenheimer/Murphy is caught off guard, thinking that Pash/Affleck and a fellow officer want to ask him questions about his lefty student Rossi Lomanitz (played by Josh Zuckerman), a victim of the communist hysteria of the 40s and 50s, who ended up "living in a hovel on the edge of a swamp [working] as a day laborer" in the early 1950s (Herken, p. 273). While he incriminated several of his associates in his HUAC testimony in 1949, Oppenheimer did not give Pash Chevalier's name. But he badly underestimated the intelligence of the security officers questioning him and thought he could get away with concocting, on the spot, what he would later call a "cock-and-bull story" (a quaint phrase that I associate with movies like *The Maltese Falcon*). A tape of the interview would come back to haunt Oppenheimer in 1954, prompting his defeatist admission to Robb that he had been "an idiot." Pash also testified in 1954. When asked about his interview a decade earlier, Pash tells Robb that he doesn't know how the story ended because he was transferred to Europe. I knew he'd been transferred. It's in my talk. He became the head of the Alsos mission to check up on the German bomb project. But I'd never stopped to ask myself why. Nolan gives the answer: Groves needed him out of his hair!

Groves got Oppenheimer to reveal Chevalier's name to him but only after he promised he'd keep that information to himself (the movie doesn't make that completely clear, I thought, but the scene with Damon and Murphy discussing this on the train is still one of my favorites). Groves probably agreed to that condition because he expected the name to be of someone he already knew to be a communist, Oppenheimer's brother Frank or one of his students. That the name was new to him put him in a difficult predicament. He didn't want to break his promise to Oppenheimer but he also realized it was a serious security breach to keep this name to himself. On the eve of the Oppenheimer hearings, in an effort to limit the damage Groves could do to their case, Strauss and FBI director Hoover made it perfectly clear to Groves that he could get into trouble himself for withholding the name of a suspected spy (Herken,p. 271). Groves probably didn't need the reminder. As Nolan made me realize and Herken (p. 112) confirms, Groves sent Pash to Europe because they knew he had compromised himself in the Chevalier affair. He did the same with John Lansdale (not in the movie) who had also interviewed Oppenheimer about the Chevalier affair. For years, I've been showing my students a picture of Pash and Lansdale in Strasbourg, where Heisenberg's colleague von Weizsäcker had set up shop. I always presented it as a complete coincidence that both of them also played an important role in the Chevalier incident. Although I had read Herken's book before, it only dawned on me watching Nolan's movie that this was no coincidence at all.

A lot more can be said about this movie but I'll limit myself to two final points, one about the storyline involving Einstein (which, as a former editor on the Einstein Papers Project, I feel I have to) and, related to that storyline, the final message with which Nolan decided to send his audience home.

I was surprised that Nolan left out the scene of Einstein (played by Tom Conti, who manages to steer clear of the usual caricature) and Szilárd writing to FDR. Maybe Nolan thought that was a little too trite (the movie does mention the letter). The first time I saw the movie, I was puzzled (and put off) by one of the Einstein scenes that Nolan did put into the movie. This is where Oppenheimer goes to see Einstein to ask him about calculations indicating that there is a small chance that an atomic bomb might ignite the atmosphere. No such visit ever took place. Einstein didn't know anything about nuclear physics: scores of physicists at Los Alamos, even some of the most junior ones, would have been in a much better position to do those calculations. Furthermore, given Einstein's perceived politics (he was enough a leftie to be on a watch-list of the FBI), this would have been a security breach comparable to his visit to ex-girlfriend Jean Tatlock (a great performance by Florence  Pugh). I warmed to the scene the second time I saw the movie, absorbing the full Einstein storyline (fictitious as it is). In between, I learned from Herken’s book  (p. 66) that the scene of Oppenheimer’s wartime visit to Einstein is not *completely* made out of whole cloth. There was a worry about igniting the atmosphere but it was about (Teller’s earliest ideas about) the H-bomb, not the A-bomb and a short calculation by the top theorist, Hans Bethe, quickly put the matter to rest. Yet, Oppenheimer was sufficiently worried about this possibility that he wanted to discuss it with one of the higher-ups. Since he felt it was too sensitive to discuss over the phone, he took a train from Berkeley—this happened before the “T[theory] section” in Los Alamos was ready—to Michigan, where he visited Nobel laureate Arthur Holly Compton, vacationing in his summer house on a lake there. So we get a glimpse of how the sausage is made by Nolan: H-bomb becomes A-bomb, lake in Michigan becomes pond in Princeton and Compton, whom the general public wouldn’t know from Adam, becomes Einstein.

When I saw the movie the second time, I also understood that the point of Oppenheimer's visit is not so much to have Einstein do the calculation but to advise Oppenheimer on what to do with its result. And Einstein's advice (like Compton's presumably) is very sensible: if there is a serious chance to ignite the atmosphere, stop and inform the Germans so that nobody destroys the world. The scene with Oppenheimer’s visit nicely sets up the scene with Oppenheimer and Einstein in Princeton after the war. This is when Strauss is offering Oppenheimer the job as Director of the Institute for Advanced Study. There is the first harbinger of the trouble ahead when Oppenheimer says he'll consider the offer instead of immediately accepting it and thanking Strauss profusely. I like the way the movie returns to that scene at the end. It nicely resolves the issue Strauss has been obsessing about, namely what Oppenheimer said to Einstein to cause Einstein to snub him. As Strauss’ aide suggests casually: maybe they weren't talking about him, maybe they had more important things to talk about. We then finally get to see the actual conversation. The conversation has two parts. One is the bit about igniting the atmosphere, where Oppenheimer tells Einstein, in effect, that the arms race is likely to make his wartime worries about destroying the world come true. The other part puts a nice spin on the scene that follows shortly thereafter, in which Oppenheimer receives the Fermi award out of the hands of Lyndon Johnson, barely a week after the assassination of John F. Kennedy, who made the decision to give Oppenheimer this award. In my talk, I used this rehabilitation of Oppenheimer to end on a positive note after the disgrace of the 1954 hearings. Nolan doesn't need it for that: he uses the much better Strauss comeuppance story for that purpose. Instead he can use the 1963 scene to show how Einstein's prediction came true: if you stick around long enough, they'll eventually pin a medal on your chest, at which point it's not really a medal for you anymore but for the generation that came after you, following in your footsteps, but leaving you in the dirt. I thought that was a nice touch. Most viewers, however, will remember this scene primarily because of the "igniting the atmosphere" threat, which Nolan used for the grand finale of his movie.

That final scene, where Oppenheimer sees the planet about to be engulfed in flames echoes an earlier one in which Borden—the former congressional aide doing Strauss' dirty work who had been a pilot in the Air Force during World War II—sees German rockets heading towards Britain and worries about the future. At the beginning of the war, while still a student at Yale, Borden had already written a grim book about the possibility of a "nuclear Pearl Harbor" (Herken, p. 194). And shortly after the war, he and some friends had placed a newspaper ad demanding that the US issue a nuclear ultimatum to the Soviets (ibid.). This ad is what first drew Senator McMahon's attention to Borden. Nolan somehow manages to pack all that information into one short scene with Borden flying his plane and seeing those German rockets. And then he creates a similar scene with Oppenheimer.

Nolan clearly wanted to send the message that we should continue to be extremely concerned about nuclear weapons. He overdoes that a bit for my taste. Nobody should lose sleep over nuclear bombs igniting the atmosphere. Nuclear armageddon is not the biggest threat facing mankind. Climate change is. I can see this with my students. None of them are worried about nuclear weapons. All of them are worried about climate change.[3] So I would have preferred Nolan to highlight a different message, which, while also part of the movie, doesn't get pride of place. And that is the danger of demonizing scientists for telling politicians things they don't want to hear. Fauci is a good example. And this stands even if it should turn out that the hard lockdown he championed did more harm than good in the end. Just as Oppenheimer remains a good example regardless of whether, in hindsight, one agrees or disagrees with his opposition to the development of the H-bomb. The most insidious current example is the demonizing of climate scientists. As the movie makes clear, there are difficult questions about what role scientists should play in societal decisions about the use of their findings. But what should be non-controversial is that scientists are not going to be punished for giving politicians and thereby society advice based on their honest assessment of the science they are asked to evaluate. That to me is the main take-away from the Oppenheimer story.

[3] A question from the audience at an early version of my post-movie talk made me realize that Nolan's message is actually broader. I incorporated that message in subsequent versions of my talk. It is conveyed by two lines that both occur twice in the movie and it is hinted at when Rabi tells Oppenheimer that he is not coming to Los Alamos: "I don't wish the culmination of three centuries of physics to be a weapon of mass destruction." The message, it seems to me, is this: the Enlightenment notion that our mastery of nature through science will solve all problems of mankind has been tempered by the sober realization that we mess with nature at our own peril (Nolan: "You can lift the rock without being ready for the snake that is revealed"). Nuclear weapons are exhibit B for this thesis; climate change is exhibit A. In both cases, the public, or large segments of it, don't seem to recognize the danger until they have seen the consequences with their own eyes (Nolan: "Theory will only take you so far").

## Bibliography

**Main sources:**

Kai Bird and Martin J. Sherwin, *American Prometheus. The Triumph and Tragedy of J. Robert Oppenheimer*. New York: Knopf, 2004.

Christopher Nolan, *Oppenheimer. The Complete Screenplay*. London: Faber, 2023

Gregg Herken, *Brotherhood of the Bomb. The Tangled Lives and Loyalties of Robert Oppenheimer, Ernest Lawrence, and Edward Teller*. New York: Henry Holt, 2002.

Jeremy Bernstein, *Oppenheimer. Portrait of an Enigma*. Chicago: Ivan R. Dee Publisher, 2004.

**Additional sources:**

David C. Cassidy, *Beyond Uncertainty. Heisenberg, Quantum Physics, and the Bomb*. New York: Bellevue Literary Press, 2009.

Jennet Conant, *Tuxedo Park. A Wall Street Tycoon and the Secret Palace of Science that Changed the Course of World War II*. New York: Simon & Schuster, 2002.

———, *Man of the Hour. James B. Conant. Warrior Scientist.* New York: Simon & Schuster, 2017.

Peter Goodchild, *Edward Teller. The Real Doctor Strangelove*. Cambridge: Harvard University Press, 2004

Priscilla J. McMillan, T*he Ruin of J. Robert Oppenheimer and the Birth of the Modern Arms Race*. New York: Viking, 2005.

Ray Monk, *Robert Oppenheimer. A Life Inside the Center*. New York: Doubleday, 2012.

Iric Nathanson, "The Oppenheimer Affair. Red Scare in Minnesota." *Minnesota History*, Spring 2007, pp. 172–186.

Michael J. Neufeld, *Von Braun. Dreamer of Space. Engineer of War*. New York: Vintage Books, 2007.

James L. Nolan, Jr., *Atomic Doctors. Conscience and Complicity at the Dawn of the Nuclear Age*. Cambridge, MA: The Belknap Press of Harvard University Press, 2020.

Richard Rhodes, *The Making of the Atomic Bomb*. New York: Simon and Schuster, 1986.

———, *Dark Sun. The Making of the Hydrogen Bomb*. New York: Simon and Schuster, 1995.

Ruth Lewin Sime, *Lise Meitner. A Life in Physics*. Berkeley: University of California Press, 1996.

Roger H. Stuewer, *The Age of Innocence. Nuclear Physics Between the First and the Second World War*. Oxford: University of Oxford Press, 2018.

Charles Thorpe, *Oppenheimer: The Tragic Intellect*. Chicago: University of Chicago Press, 2006.